# Network visualization techniques for story charting

## The case of a book in Portuguese

Joao T. Aparicio
INESC-ID, Instituto Superior Tecnico,
Universidade de Lisboa
Lisbon, Portugal
joao.aparicio@tecnico.ulisboa.pt

Andreas Karatsolis
Department of Comparative Media Studies and Writing
Massachusetts Institute of Technology
Cambridge, MA, USA
karatsol@mit.edu

Carlos J. Costa
Advance/CSG, ISEG (Lisbon School of Economics & Management)
Universidade de Lisboa
Lisbon, Portugal
cjcosta@iseg.ulisboa.pt

*Abstract* — **Visualization techniques have been widely used to analyze various data types, including text. This paper proposes an approach to analyze a controversial text in Portuguese by applying graph visualization techniques. Specifically, we use a story charting technique that transforms the text into a graph. Each node represents a character or main entities, and each edge represents the interactions between characters. We also present several visualization techniques to gain insights into the story's structure, relationships between the characters, the most important events, and how some key terms are used throughout the book. By using this approach, we can effectively reveal complex patterns and relationships that may not be easily discernible from reading the text. Finally, we discuss the potential applications of our technique in Literary Studies and other fields.**

***Keywords - text mining, data visualization, network visualization, Voyant Tools.***

### I. Introduction

As the number of written reports produced in corporate and noncommercial contexts increases, the need to summarize their content becomes even more significant. While data summarization is typically performed numerically or graphically, text summarization is breaking down long documents into fewer paragraphs or sentences. The procedure extracts essential information, while ensuring the sections preserve the original meaning. Many approaches to text summarization have been implemented [8], and El-Kassas et. Al [3] have presented an extensive analysis and typification of summarization approaches. As it is also relevant to present results in a way that makes it easy to see and interpret the results obtained, text visualization has been a field of significant interest [6,9]. Overall, such approaches fall into the domain of text mining.

For the purposes of this project, our purpose was to visually summarize the essential facts, events and participants graphically from a controversial book in Portuguese describing the collapse of the BES Bank and its significant impact for the Portuguese economy. The main question we had to address is : how can one represent the summary of a document with a story graphically? In order to answer this, we tackled the following sub-questions: What are the most frequent words? What is the evolution of occurrence of names (protagonists, roles, activities/events)? How are the words related?

To answer our research questions, we used data analysis techniques, including network science [26,31] and cluster analysis, supported by data visualization techniques.

### II. Background

Some questions are complex and can be viewed from different perspectives. Therefore, it is essential to summarize and show patterns, but also to preserve the original meaning of the text. Visualization is an important part of interpreting texts. Liu [15] found that readers need to familiarize themselves with various ways to make sense of and read visual images. Guo-dong [16] found that visualization is a primitive stage and fundamental form of deverbalization in the course of interpreting, and it is essentially a dynamic procession of mental spaces triggered by language. Fortuna [17] found that visualization of text document corpus can be quite useful for complex tasks. These findings suggest that visualization is an important tool for interpreting texts. Visualization can help readers understand texts' meaning by extracting important concepts and ideas [27-28 , 32].

Visualizations can also be used for objective analysis of a text. Cao [18] provides an overview of text visualization techniques, Wise [19] found that spatializing text content can be used to enhance visual browsing and analysis, Goffin [20] found that word-scale visualizations can be used to provide additional information about the text, and Morse [21] found that visualizations can be used to support users in information-seeking environments. Following this literature on objective analysis of text, we now focus on how visualization techniques can be useful for summarizing, exploring, and presenting stories.

For comprehending vast quantities of written material, text representation is essential in natural language processing (NLP). Text visualization techniques and methods range from basic methods like word clouds, to more complicated methods like subject modeling and network analysis. Wordle, Voyant, Gephi, Python's Natural Language Toolkit (NLTK), and Scikit-learn modules are among the most common text rendering tools. [1][4]

Word clouds are a quick and easy method to display text data. They show the most frequently occurring terms in a text collection, with the height of each word showing its prevalence in the text. While word clouds are simple to make and can provide a fast summary of the most prevalent terms in a text, they lack in-depth analysis. Other techniques are needed for more sophisticated text data processing. Furthermore, word clouds are readily altered by altering the typeface height or hue, which may result in skewed readings [22]. Topic modeling is a common text visualization technique that includes categorizing words and sentences in a collection into themes. Each subject symbolizes a group of connected words or sentences, with the intention of identifying the text's central ideas or concepts. Latent Dirichlet Allocation (LDA) is a common subject modeling method [23]. LDA is a statistical model that assumes that each text is a collection of topics, with each topic being a probability distribution over words. Sentiment analysis is a text visualization technique that includes determining the emotional tone of a small amount of text. Sentiment analysis can determine the mood of a text collection as a whole or pinpoint the attitude of particular words or sentences. Sentiment analysis approaches include lexicon-based methods [25], machine learning methods [4], and mixed methods [14].

Network analysis displays text in which the connections between words or sentences are represented as a network. Each component in a text network symbolizes a word or sentence, and the lines reflect the connections between them. Network analysis can be used to find the most significant terms or sentences in a text collection and the text's fundamental structure. TextRank [13] is a famous network analysis method that employs a version of PageRank to find the most significant words or sentences in a text. Word embeddings display text in which words are represented as vectors in a high-dimensional environment. Word embeddings can be used to determine meaning connections between words, such as compounds and antonyms, as well as tasks such as word comparison and mood analysis. Word2Vec [12] is a famous word-embedding method that employs a neural network to acquire word vector forms.

The process of depicting a text collection as a network, with nodes representing words or ideas and lines showing the connections between them, is known as network analysis. Network analysis can reveal information about the connections between ideas in a text collection, such as which concepts are most closely linked or which concepts are essential to the corpus' general structure [11]. Co-occurrence analysis is a common method for text network analysis, in which terms that appear together frequently are linked by an edge in the network [6].

Research challenges in visualizing text using networks include finding a way to group similar messages into clusters automatically and using semantic analysis to manage the representation's complexity [9-10].

A way to manage this complexity is by preselecting the elements for the analysis of the graph. To do that, we may look at a story and its main elements, the characters, by using named entity recognition and matching the selection based on both the number of interactions and the number of references. Then graph clustering techniques can be applied to group characters and sets of interactions.

## III. METHODOLOGY

In order to tackle the problem of visual summarization, the following steps were performed:

- Word clouds were generated to allow the selection of the most frequent characters, events, and subjects.
- The evolution of the frequency of specific words referenced throughout the text was analyzed.
- The co-occurrence of main words, which allows for the creation of a network of entities, was examined.
- The topological features of the network created were identified, in order to determine clusters of concepts and participants in the narrative in groups.

To apply the network representation and analysis, we use a name entity recognition method based on Bert, pre-trained for the Portuguese language [10]. First, we identified the proper Nouns present in the text. We then selected the ones most relevant to the analysis, based on context and calculated phasic co-occurrence of names. We finally applied a modularity maximization method for building the network communities of characters in the book. This enables us to understand who the main characters are, how they interact, how they group, what are the main groups and who are the most influential characters for connecting different sets of characters. This systematic approach effectively enables us to create a character social network based on any written text in Portuguese. For task management we used the POST-DS methodology for this project [29-30].

## IV. RESULTS

We started to analyze the most frequent words, after removing Portuguese stopwords. The result was displayed in a word cloud. It is interesting verifying that BES and Salgado were two of the most frequent words.

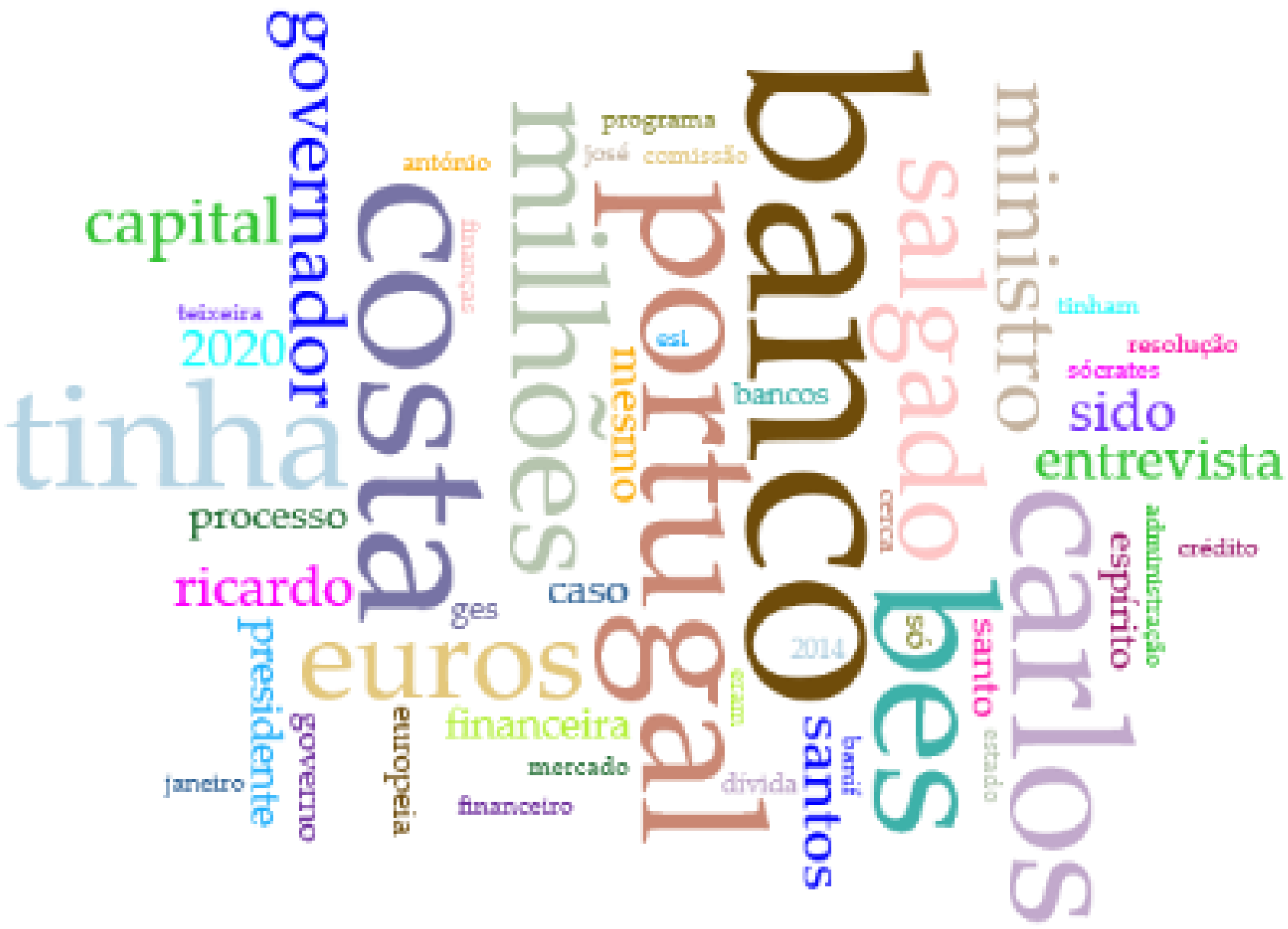

Figure 1. The .main terms.

In the following figures, we present the evolution of references of characters along the text. The main character is the former governor, Carlos Costa (not the author of this paper). However, this story also has some other crucial characters: the prime minister and the antagonists. The prime ministers during this period are José Socrates, Pedro Passos Coelho, and António Costa. The antagonists are Ricardo Salgado and Isabel dos Santos, protagonists of activities subject to decisions from Banco de Portugal's governor.

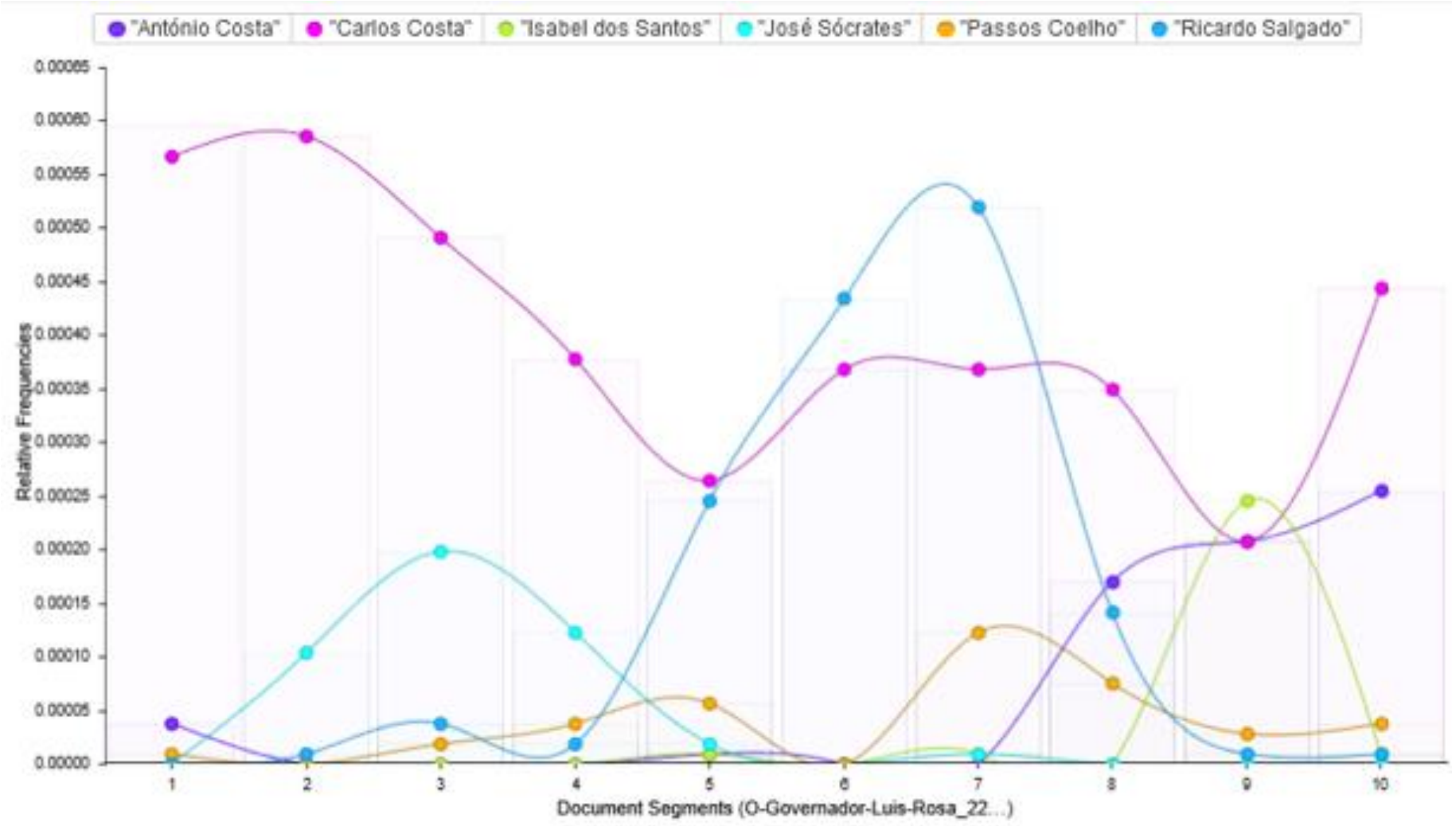

Figure 2. The main character, prime ministers, and other main participants

The author of the book refers to other governors in addition to Carlos Costa. The previous governor was Vitor Constâncio and the next one Mário Centeno.

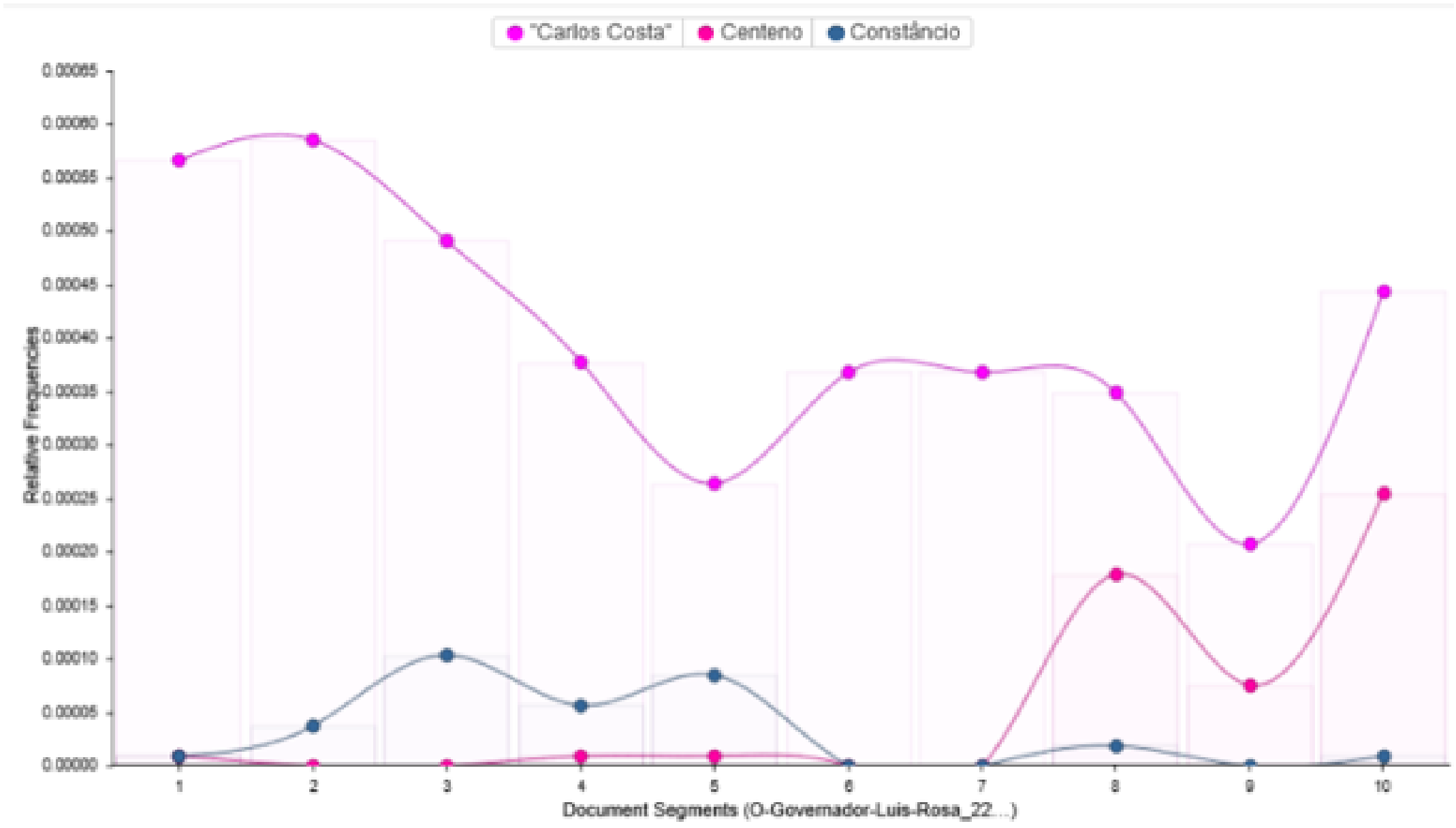

Figure 3. The Governors.

The most critical events are related to the two resolved banks: BANIF and BES. Another important event is the Troika. Troika appears in the beginning as the most frequent among those words. Then, the most important part of the document relates to BES and, finally, BANIF.

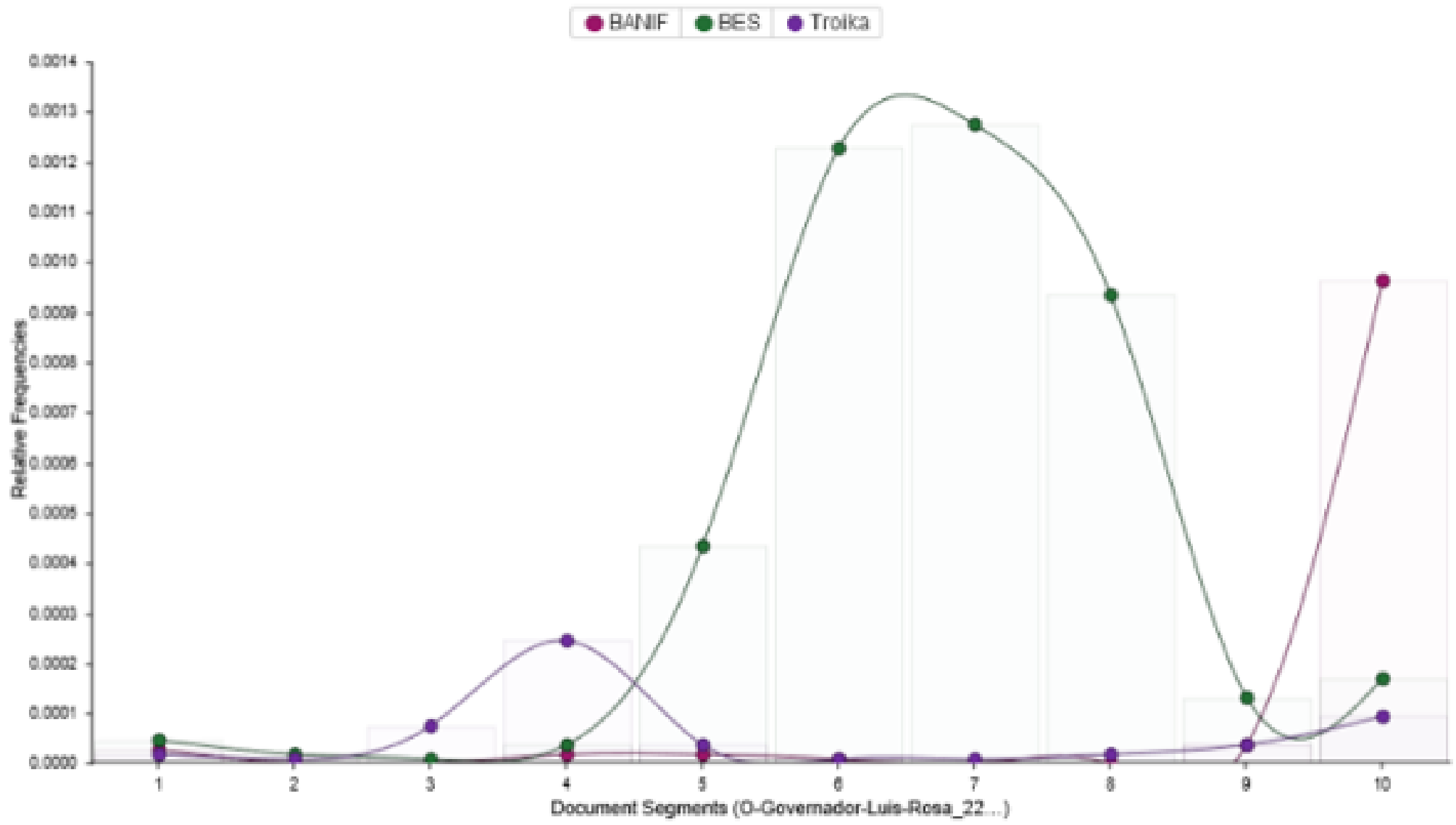

Figure 4. References to BANIF, BES and Troika.

Some prevalent words in the text are case (*casos*), process (*processo*), crisis (crise), and pressure (*pressão*). The text starts with a particular emphasis on the "crisis" and ends with a special focus on the word "process". The process refers typically to the legal process.

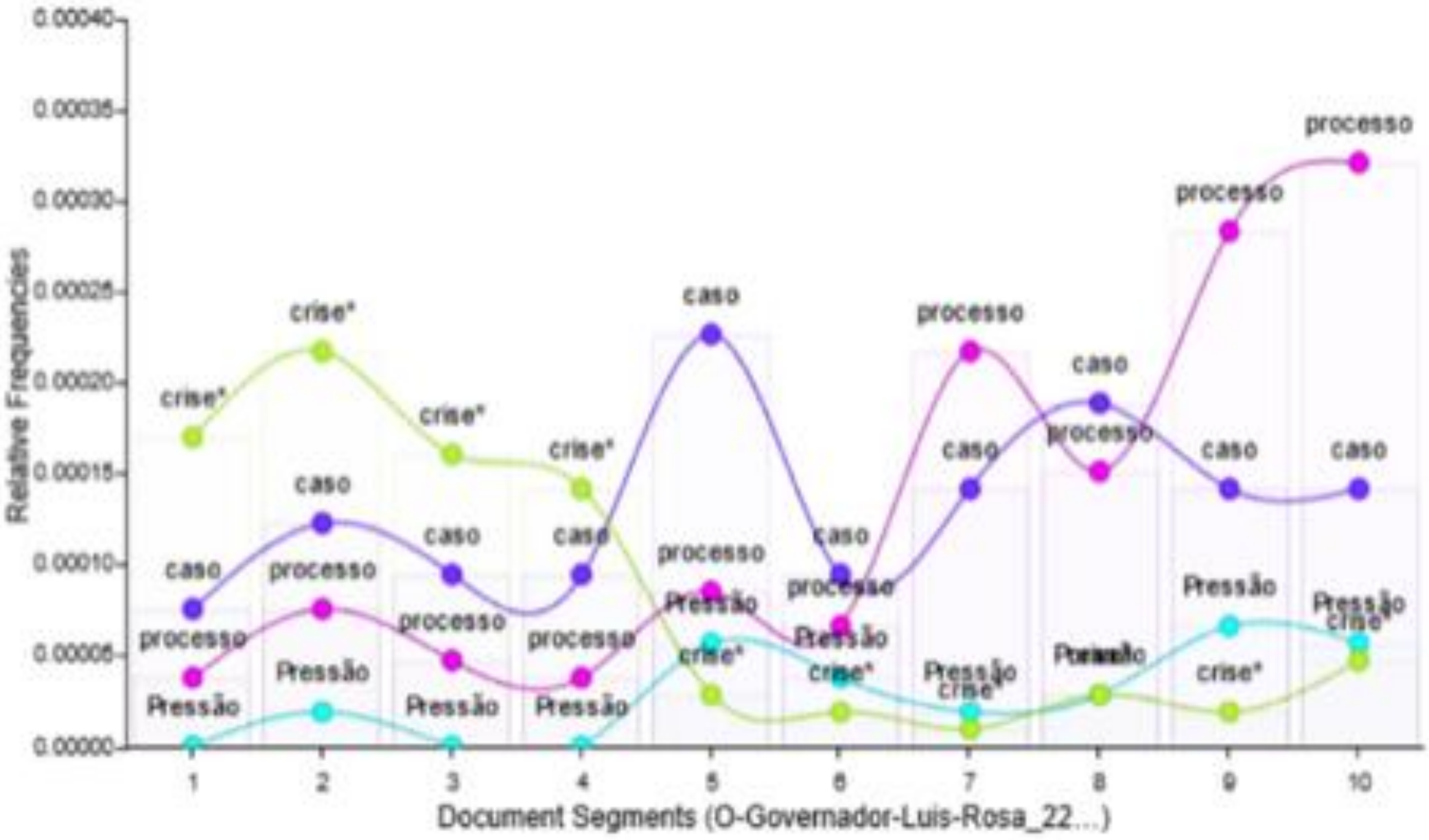

Figure 5. Case, process, crisis, and pressure

Then, we analyze how the words are related. Network analysis allows for analyzing the relationship between terms.

With the named entity recognition and co-occurrence data we generated, we built the graph in Figure 6, where node size is the frequency of the noun in the text and edge thickness is the number of co-occurrences. From there, we can see that BES and Carlos Costa are the most prominent entities in the text we are studying.

Using network metrics, we can extract additional information from the generated graph. Namely, we can use the betweenness centrality that measures how important a node is at connecting the network (i.e., the ratio of the shortest paths that go through that node). We can also calculate the network communities to understand the extent of the connection between different network elements. (see Figure 7).

As we can see, betweenness centrality gives us different semantic information on the context. Previously, we saw the importance of the entities based on their occurrence in the text. However, we now see entities' importance in connecting other entities in this case. For instance, we can see that Ricard Salgado and Governo (Government) are much more important in connecting the case.

The groupings generated can show us three different communities. This number was obtained by optimizing network modularity using the Louvain algorithm, meaning there was no definition of the number of communities beforehand. We can clearly see one community associated with central banking (light green), another related to BES (dark green), and another related to the Portuguese government (grey). However, it is interesting that entities like António Costa and Mário Centeno are not on the grey cluster.

Factor analysis and cluster analysis may be used respectively to summarize information by reducing the number of variables and aggregating the data (in this case, words or terms) in groups (or clusters). There is the option to use PCA (principal component analysis) or correspondence analysis for factor analysis. In the second part, we may use several cluster analysis techniques for creating groups. Correspondence analysis is a multivariate statistical technique that can be used to display or summarize data sets in a smaller number of dimensions. Like in PCA, in correspondence analysis the goal is to reveal any structure hidden in the multivariate sample. PCA and cluster analysis allowed us to identify three factors (in Figures 8 and 9) and 3 clusters with three factors, a total variance explained of 84.2%. We can distinguish a cluster related to the relationship with outside entities (purple), a cluster related to the internal management of the bank of Portugal (green), and a group that only includes Ricardo Salgado, BES, and BESA.

The use of these techniques allowed us to determine that the main protagonist (Carlos Costa) was referred to uniformly over time. The prime ministers were referred to sequentially in chronological terms (Socrates, Passos Coelho, and Costa). It was also possible to identify the most important events and how some powerful words are referred to through the text. The same happened with the cases/processes: Crise, BES, and Banif. The principal suspects to the case were also mentioned chronologically: Salgado and Isabel dos Santos.

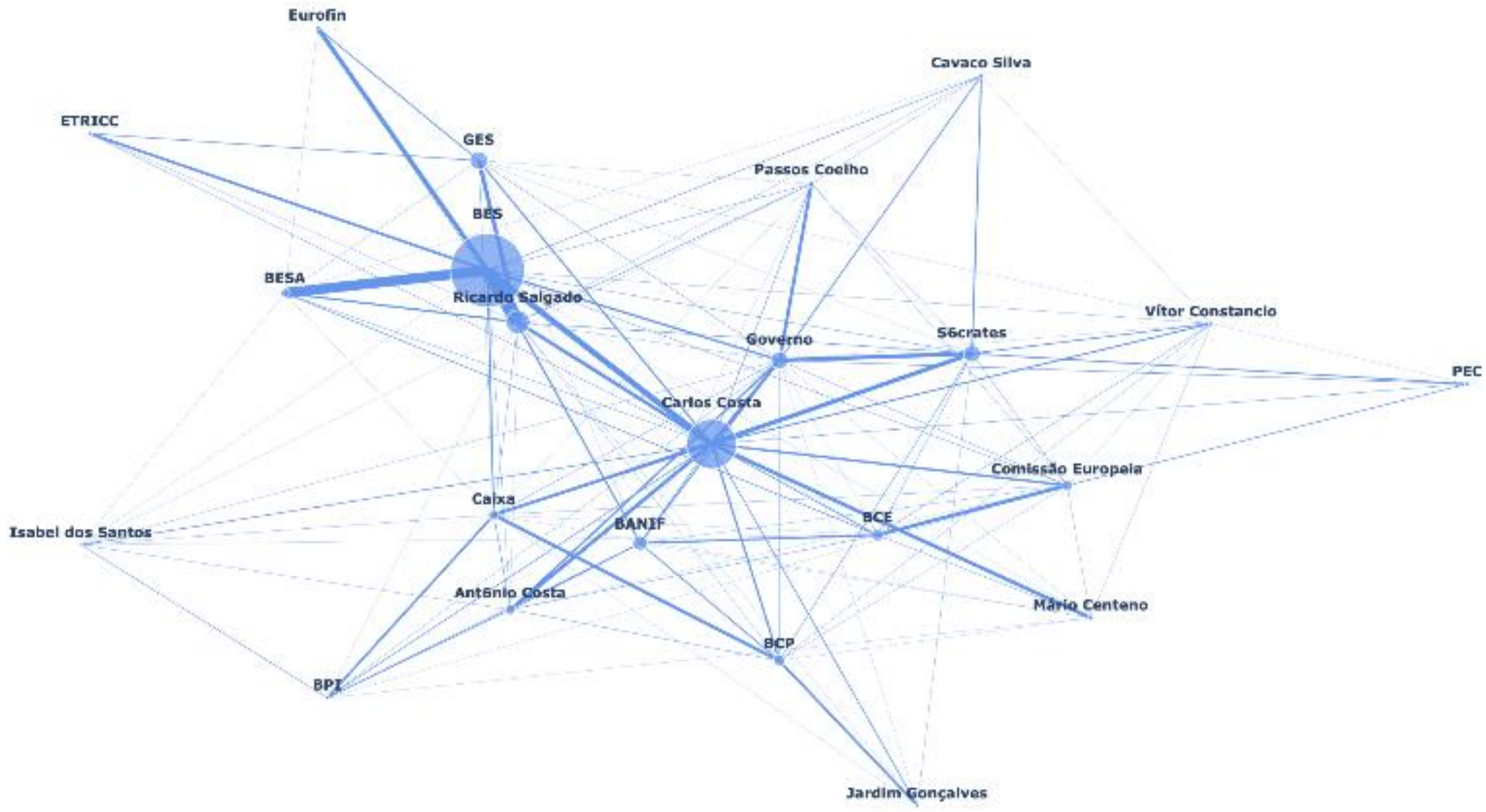

Figure 6. A network of co-occurrence of selected terms.

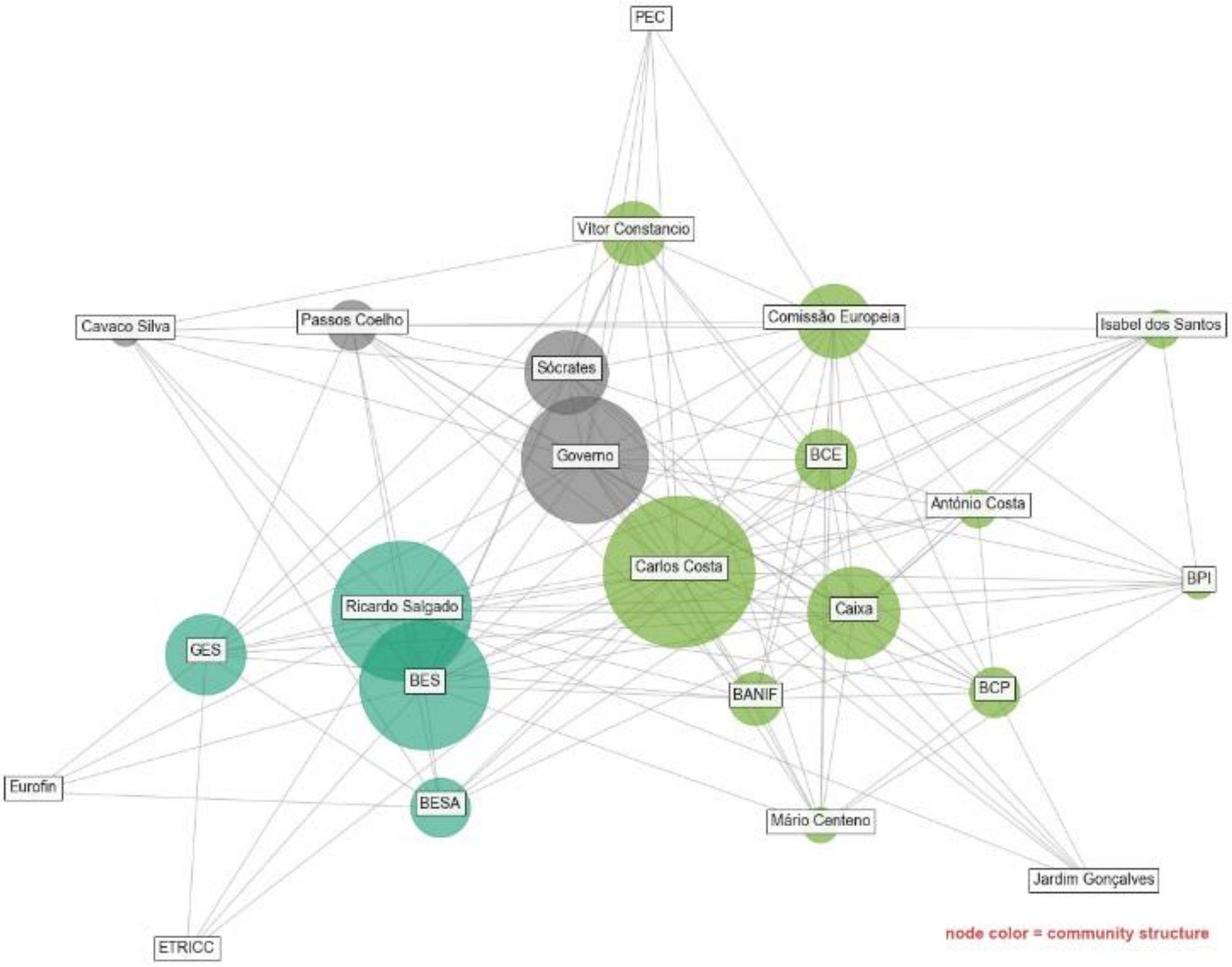

Figure 7. Network communities and betweenness centrality (node size) of co-occurrence of selected terms

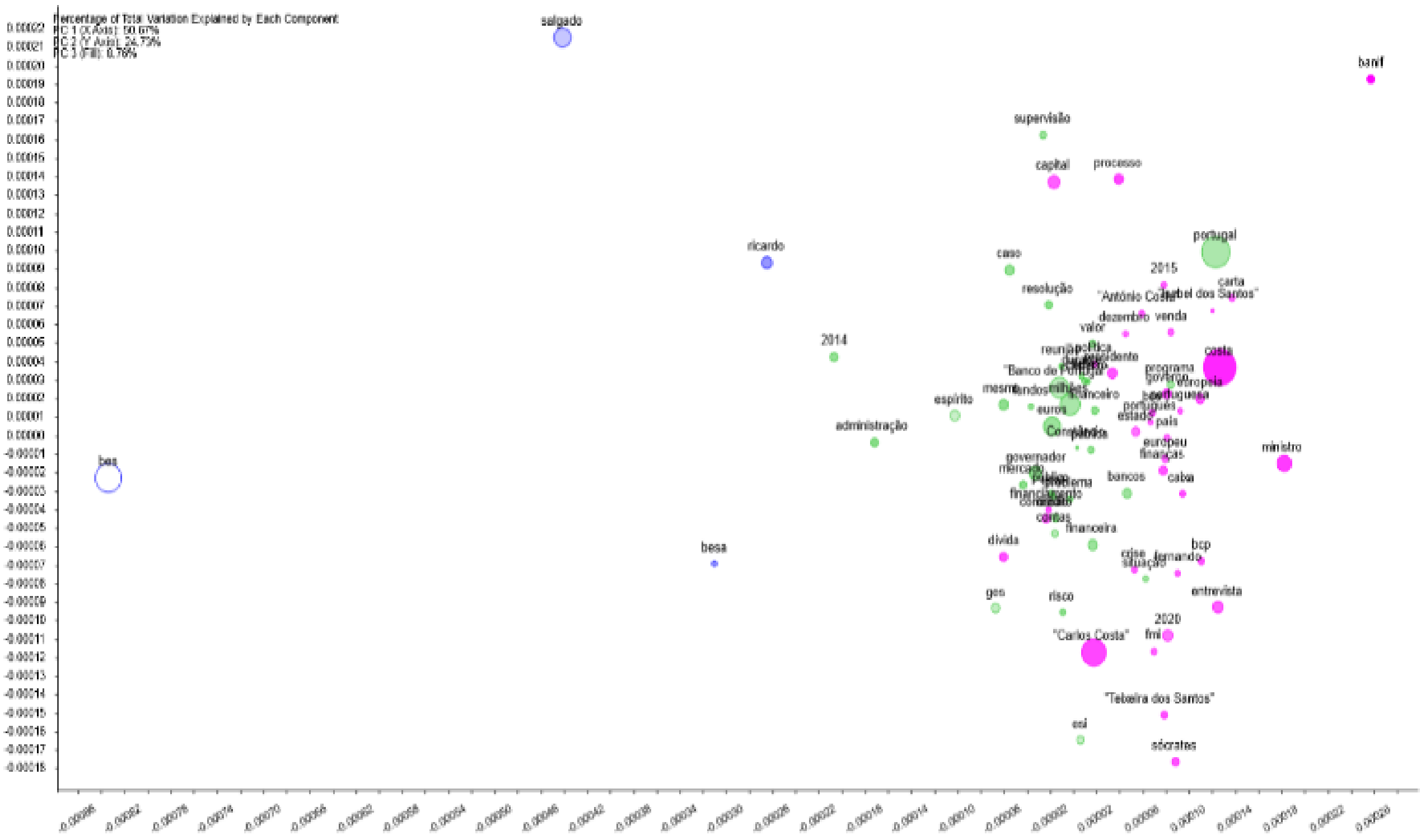

Figure 8. PCA with three factors and three clusters

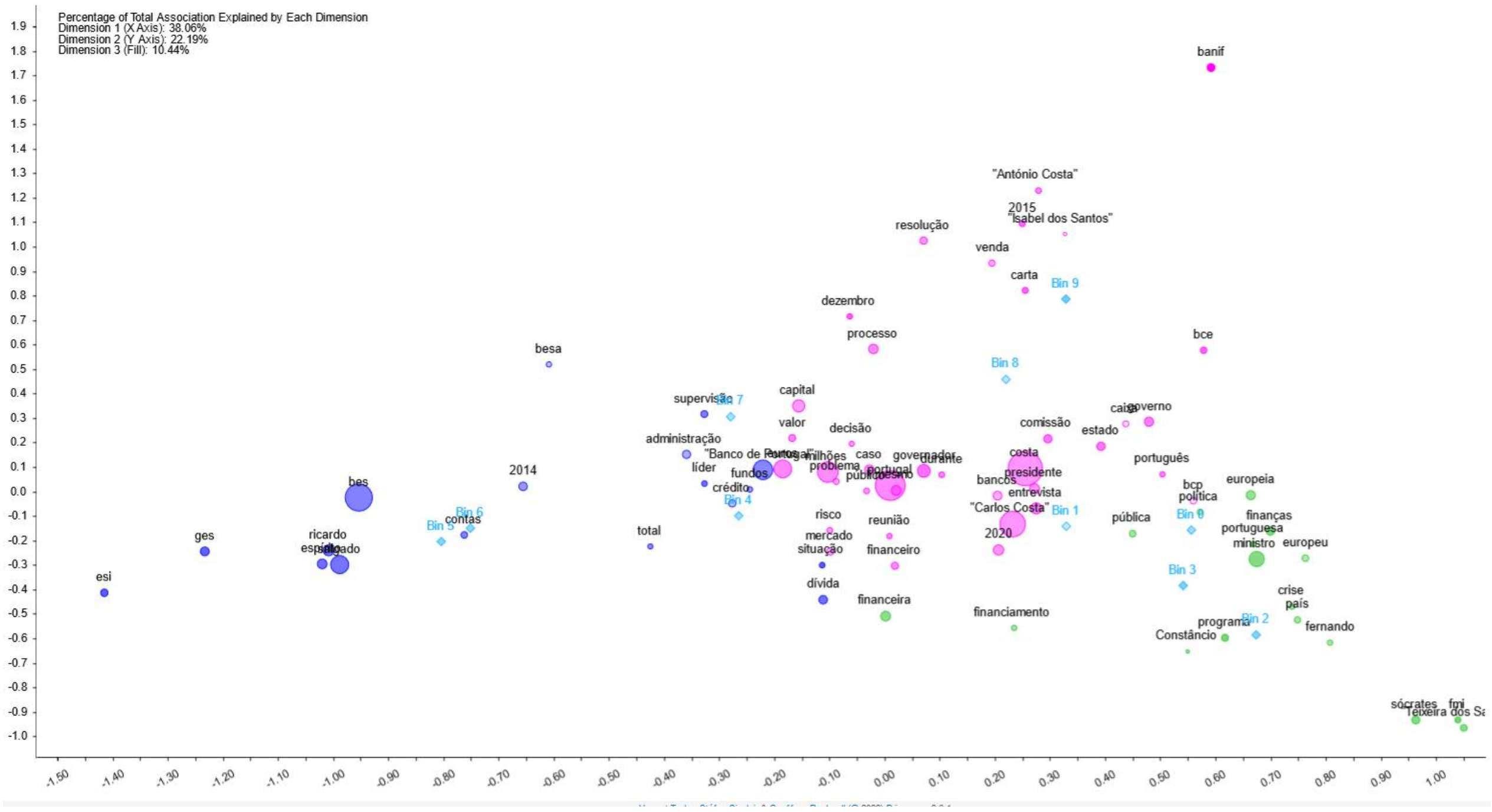

Figure 9. Correspondence analysis.

The research presented here primarily aims to understand the content of a polemic book using objective methods. Specifically, the purpose was to identify the main actors referred throughout the book, the main terms, and how those are related. Future related work could include performing a more profound analysis between events and actors. We suggest using natural language processing with semantic approaches to tackle this problem.

## V. CONCLUSIONS

Our research allowed us to identify an approach that summarizes a large text using visual representations. We started with a straightforward approach like a word cloud, but the need for a more sophisticated approach allowed us to gain greater insights about the text. This method allows us to bring network visualizations through character social networks to any long text using machine learning pre-trained models for the Portuguese language. The way the characters are selected using reference and interactions and grouped through network community findings helps us tackle the problem of complex text visualizations. The potential implications of this study include a better understanding of the narrative structure of controversial texts and the ability to identify underlying themes and social networks within the text. The approach could have applications in literary studies and fields such as social network analysis and psychology. In the future, the exploration of different network visualization layouts for information representation may be of value to help the user easily grasp the complex net of entities and relationships in highly complex texts.

## ACKNOWLEDGMENTS (HEADING 5)

We gratefully acknowledge financial support from FCT -Fundação para a Ciência e a Tecnologia (Portugal), national funding through research grant UIDB/04521/2020. This work is also supported by national funds through PhD grant (UI/BD/153587/2022) supported by FCT.